\documentclass[10pt,journal=jpclcd,manuscript=article]{achemso}

\usepackage[utf8]{inputenc} 
\usepackage[T1]{fontenc}    
\usepackage{hyperref}       
\usepackage{url}            
\usepackage{booktabs}       
\usepackage{amsfonts}       
\usepackage{amsmath}        
\usepackage{amssymb}        
\usepackage{nicefrac}       
\usepackage{microtype}      
\usepackage{lipsum}
\usepackage[]{graphicx}       
\usepackage{gnuplot-lua-tikz}
\usepackage{caption}
\usepackage{subcaption}     
\usepackage{float}          
\usepackage{cleveref}       
\usepackage{multicol}       
\usepackage{geometry}
\usepackage{color}

\title{Statistically Unbiased Free Energy Estimates from Biased Simulations}

\author{Matteo Carli}
\affiliation{SISSA, Via Bonomea 265, 34136 Trieste, Italy}
\author{Alessandro Laio}
\affiliation{SISSA, Via Bonomea 265, 34136 Trieste, Italy}\email{laio@sissa.it}

\begin{tocentry}
\includegraphics[width=2in]{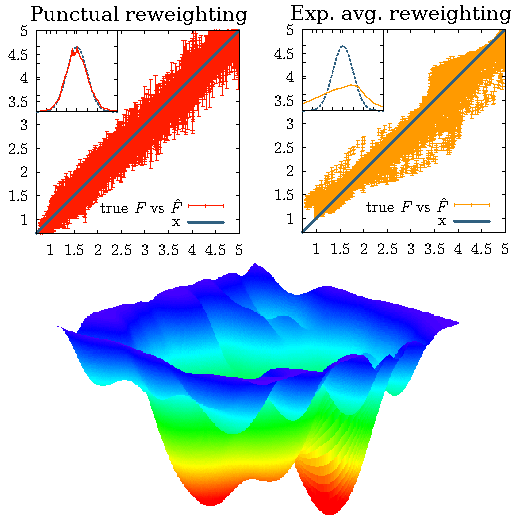}
\end{tocentry}


\SectionNumbersOn

\begin{document}
\maketitle

\begin{abstract}
\noindent Estimating the free energy in molecular simulation requires, implicitly or explicitly, counting how many times the system is observed in a finite region. If the simulation is biased by an external potential, the weight of the configurations within the region can vary significantly, and this can make the estimate numerically unstable. 
We introduce  an approach to estimate the free energy as a simultaneous function of several collective variables starting from data generated in a statically-biased simulation. 
The approach exploits the property of a free energy estimator recently introduced by us \cite{Rodriguez2018}, which provides by construction the estimate in a region of infinitely small size. We show that this property allows removing  the effect of the external bias in a simple and rigorous manner.  The approach is validated on model systems for which the free energy is known analytically and on a small peptide for which the ground truth free energy is estimated  in an independent unbiased run. In both cases the free energy obtained with our approach is an unbiased estimator of the ground-truth free energy, with an error whose magnitude is also predicted by the model. 
\end{abstract}

\section{Introduction}
A central problem in atomistic simulations is the study of systems exhibiting a complex and multidimensional free energy landscape. This landscape is typically estimated as a function of a set of collective variables (CVs), which are explicit functions of all the coordinates of the system\cite{fiorin2013using}.
In many practical applications, the sampling of all the relevant states of the system is artificially enhanced using a wide range of methods (see \cite{bernardi2015enhanced} for a review). The prototype of many of these methods is Umbrella Sampling\cite{TorrieG.MValleau1977}, in which an external bias potential is added to the potential energy with the aim of accelerating the transitions between the local free energy minima and explore all the relevant metastable states of the system. We shall focus here on this specific approach.

During the simulation, the bias $B(\mathbf{x})$ is added to the potential energy $U(\mathbf{x})$. This bias, is a function of a (possibly multidimensional) CV $\mathbf{s}(\mathbf{x})$, namely $B(\mathbf{x}) \equiv B(\mathbf{s}(\mathbf{x}))$. In the most general case, one is willing to estimate the free energy as a function of another (also possibly multidimensional) CV $\mathbf{\sigma}(x)$.The unbiased free energy  for a specific value $\tilde{\mathbf{\sigma}}$ can be estimated as:
\begin{equation}
    F(\tilde{\mathbf{\sigma}})  = -\beta^{-1} \log \int \rho^B(\mathbf{x}) \,e^{\, \beta  B(\mathbf{s}(\mathbf{x}))} \, \delta(\tilde{\mathbf{\sigma}}-\mathbf{\sigma}(\mathbf{x})) \, \mathrm{d}\mathbf{x} \; + f^B
\label{eq:F-rho_biased-explicit_dimred}
\end{equation}
where $\rho^B(\mathbf{x}) = Z_B^{-1} e^{- \beta  ( V(\mathbf{x}) + B(\mathbf{s}(\mathbf{x})))}$ is the biased probability distribution, $Z_B$ is the biased canonical partition function and $f^B$ is an additive constant. In ordinary Umbrella Sampling the biasing CV is an explicit function of the the variables $\sigma$, namely $\mathbf{s}(\mathbf{x}) \equiv \mathbf{s}(\mathbf{\sigma}(\mathbf{x}))$. In this case eq \ref{eq:F-rho_biased-explicit_dimred} takes the simple known form $F(\tilde{\mathbf{\sigma}}) = F^B(\tilde{\mathbf{\sigma}})-B(\mathbf{s}(\tilde{\mathbf{\sigma}}))$ where $F^B(\tilde{\mathbf{\sigma}})=-\beta^{-1}\log \int \rho^B(\mathbf{x})\,  \delta(\tilde{\mathbf{\sigma}}-\mathbf{\sigma}(\mathbf{x}))\, \mathrm{d}\mathbf{x}$. If $\mathbf{s}(\mathbf{x})$ is a generic function of the coordinates, instead, the exponential factor cannot be brought out of the integral and there is no easy manner to estimate the unbiased probability density $\rho(\mathbf{\sigma})$ from $\rho^B(\mathbf{\sigma})$. 

Both in the case in which $\mathbf{s}$ is a function $\mathbf{\sigma}$ or not, what is generally done in literature\cite{Rosenblatt1956,Parzen1962OnMode,silverman1986density,Altman1992Nonparam,Kumar1992wham,Bartels1997AdaptiveUmbrella,Shirts2008,Tan2012binless,zhang2019uwham} is to relax the delta function in eq \ref{eq:F-rho_biased-explicit_dimred} and instead use a kernel $K_{\Delta}$ that converges to it only when the limit to zero is taken on its scale parameter $\Delta$ (also called smoothing parameter\cite{Park1990ComparisonSelectors}). In other words, instead of $F$ in eq \ref{eq:F-rho_biased-explicit_dimred}, the following quantity $F_K$ is computed:
\begin{equation}
    F_K(\tilde{\mathbf{\sigma}}) := -\beta^{-1} \log \int \rho^B(\mathbf{x}) e^{\, \beta  B(\mathbf{s}(\mathbf{x}))} K_{\Delta}(\tilde{\mathbf{\sigma}}-\mathbf{\sigma}(\mathbf{x})) \mathrm{d}\mathbf{x}
\label{eq:F_bias-Kernel}
\end{equation}
so that, by replacing the true biased density $\rho^B(\mathbf{x})$ by its sample estimator $\hat{\rho}^B(\mathbf{x}) = \frac{1}{N} \, \sum_{j=1}^N \, \delta(\mathbf{x}-\mathbf{x}_j)$ one obtains
\begin{equation}
    \hat{F}_K(\tilde{\mathbf{\sigma}}) \sim -\beta^{-1} \log \sum_{j=1}^N e^{\, \beta  B(\mathbf{s}(\mathbf{x}_i))} \, K_{\Delta}(\tilde{\mathbf{\sigma}}-\mathbf{\sigma}(\mathbf{x}_j))
    \label{eq:F_hat_bias-Kernel}
\end{equation}
where $\hat{F}_K$ denotes the estimator of $F_K$ in eq \ref{eq:F_bias-Kernel} and is defined up to an additive constant. One simple example of such kernels $K$ is the characteristic function $\chi_I(|\tilde{\mathbf{\sigma}}_b-\mathbf{\sigma}|)$ of the interval $I=[0,\Delta/2]$, which is the case of a histogram of bin size $\Delta$ centered around the bin centers $\{\tilde{\mathbf{\sigma}}_b{\}}_b$. The expression in eq \ref{eq:F_bias-Kernel} however also applies to other kernel methods or, with a point-dependent value of $\Delta$, to the \textit{k}-Nearest Neighbour estimator (\textit{k}-NN)\cite{Fix1951DiscriminatoryProperties,LoftsgaardenD.O.Quesenberry1965AFunction,Mack1979MultivariateEstimates}. The common feature of these estimators is that they are not punctual, namely they provide an estimate of the free energy on a finite-size region. Thus, in the limit $n \rightarrow \infty$ they converge to $F_K$ in eq \ref{eq:F_bias-Kernel} but not to $F$ in eq \ref{eq:F-rho_biased-explicit_dimred}. Over this neighbourhood the value of $e^{\, \beta B(\mathbf{s})}$ can be largely fluctuating, making the estimators ill-behaved\cite{Kobrak2003Error,Shirts2005}. The estimators $\hat{F}_K$ only converge to $F$ asymptotically, namely in the limit when both $\Delta \rightarrow 0$ and $n \rightarrow \infty$. However, even when big, $n$ will always be finite, so a finite parameter $\Delta$ may be required in order for the estimators $\hat{F}_K$ to be statistically meaningful. This problem becomes more and more severe in high dimension, due to the curse of dimensionality\cite{beyer1999nearest} and can happen even in the trivial case $\mathbf{\sigma} = \mathbf{s}$, if $\mathbf{s}$ is multidimensional or the sample is too small. 

We here show that this problem can be circumvented by exploiting  a \emph{punctual} free energy estimator, i.e. one in which the limit for $\Delta \rightarrow 0$ is implicitly taken. Under these conditions, it is not anymore necessary to take the average of the bias factors $e^{\, \beta B(\mathbf{s}(\mathbf{x}))}$ over the neighbourhood of $\mathbf{\sigma}(\mathbf{x}_i)$ set by a finite $\Delta$: the reweighting involves only the punctual value of the bias applied when generating a datapoint $i$.

An estimator with such punctual property was introduced by us in ref \cite{Rodriguez2018}. This procedure, which we called Point-Adaptive k-nearest neighbour free energy estimator (PA\textit{k})  allows estimating the free energy in a very large dimensional  space, ideally including all the relevant degrees of freedom of the system (for example, the coordinates of all the heavy atoms of a solute). The advantage of PA\textit{k} is that, for all points $\{\mathbf{\sigma}_i{\}}_i$ in a sample, it provides an unbiased estimate of the free energy $F(\mathbf{\sigma}_i)$ in eq \ref{eq:F-rho_biased-explicit_dimred} rather than the one in eq \ref{eq:F_bias-Kernel}.

As we will show, this allows removing the effect of the bias in a simple, numerically robust and theoretically well-founded manner, also in the case in which the CV on which the bias is applied is not an explicit function of the variables $\mathbf{\sigma}$.
\section{Methods}
\subsection{The PA\textit{k} approach}
\label{ssec:PAk}
PA\textit{k} is a generalization of the kNN density estimator in which the optimal value of $k$ is determined independently for each datapoint by an unsupervised, non-parametric procedure. The approach works also if the dimensionality $D$ of the space $\Sigma$ of the coordinates $\mathbf{\sigma}$ is very large (for example the position of all the $C_{\alpha}$ carbons in a protein \cite{Sormani2019,Carli2020CandidateSimulations}). What makes the estimate computationally feasible is the key assumption that, due to chemical restraints, $\rho(\mathbf{\sigma})$ has a support on a manifold $\Omega \subset \Sigma$, whose intrinsic dimension $d$ is significantly smaller than $D$. We here assume for simplicity that $d$ is constant over the whole dataset and estimate it with the  TWO-NN approach\cite{Facco2017}\footnote{Notice the PA\textit{k} approach is valid also in the case of no explicit dimensional reduction $\mathbf{\sigma}(\mathbf{x}):= \mathbf{x}$, i.e. when $\Sigma \equiv \mathbb{R}^N$. In this case the free energy $F_i$ is equivalent to the potential energy of the configuration $\mathbf{x}_i$ entering the canonical Boltzmann factor. Even in this case, however, it is still appropriate to refer to this quantity as "free energy", since $\rho(\mathbf{\sigma})$ is estimated in a space of dimension $d \leq D$.}. Another important assumption of PA\textit{k} is that, at least in a neighbourhood where the density of points is approximately constant, $\Omega$ is well approximated by its tangent hyperplane $T_{\Omega}(\mathbf{x})$. Therefore, around every point $\mathbf{x}_i$ one can measure distances in the Euclidean $\mathbb{R}^D$ metric, which for points lying on $T_{\Omega}(\mathbf{x}_i)$ is equivalent to the corresponding $\mathbb{R}^d$ restriction. We indicate by $r_{i,k}$ the Euclidean distance between the $\mathbf{x}_i$ and its $k$-th-nearest neighbour. The volume of the shell between neighbours $l-1$ and $l$ of a point $i$ is given by $\nu_{i,l} = \omega_d (r^d_{i,l}-r^d_{i,l-1})$, where $\omega_d$ is the volume of the $d$-sphere of unitary radius and $r_{i,0}$ is conventionally set to $0$ (hence, the volume of the $d$-sphere enclosed between a point $i$ and its $k$-nearest neighbour is $\sum^k_{l=1} \nu_{i,l} = \omega_d \,r^d_{i,k}$). Using these data, one first finds, by a statistical test performed independently for each point $i$, the maximum value of $k$ for which the volumes $\nu$ can be considered as drawn from a probability distribution with an approximately constant density. We call this value $\hat{k}_i$ (see ref \cite{Rodriguez2018} for an explicit definition of the test). For each point $i$ one then estimates the free energy $F_i = F(\mathbf{\sigma}(\mathbf{x}_i))$ by maximising the likelihood of a model in which the free energy is assumed to depend linearly on the the neighbourhood order $l$:
\begin{equation}
\hat{F}_i \, := \, \underset{F}{\operatorname{argmax}} \; \max_{a} \; \sum_{l=1}^{\hat{k}_i} \log(e^{-F+al}e^{-\exp(-F+al)\nu_{i,l}})
\label{eq:F_PAk}
\end{equation}
For each $i$, this maximisation is equivalent to the solution of a log-linear regression model\cite{Nelder1972GeneralizedAuthor} in which the $l$ observed responses are the random variables $\mathbf{Y}^i=\{\frac{1}{\nu_{i,l}}{\}}_l$ distributed exponentially with expected value $\langle Y^i_l \rangle = e^{-F_i + al}$. Therefore, $F_i$ is the intercept of such model and its  value is equivalent to taking the limit $l \rightarrow 0$ or, analogously, $\Delta \rightarrow 0$. The maximum likelihood approach also allows estimating the uncertainty of $\hat{F}_i$, which is given by $\varepsilon_i = \sqrt{\frac{4 \hat{k}_i + 2}{(\hat{k}_i - 1)\hat{k}_i}}$. Tested on various systems, PA\textit{k} has been proven an unbiased punctual free energy estimator, outperforming other non-parametric and unsupervised methods, especially in high dimensionality. The point-adaptive selection of the neighbourhood applied by PA\textit{k} surely plays a role in achieving this task. However, we believe the extrapolation feature of the model, described above, is crucial for obtaining a punctual estimate of the free energy.

\begin{figure*}[!ht]
\begin{minipage}{.25\textwidth}
  \vspace{-1.55\baselineskip}
  \begin{subfigure}{\textwidth}
    \centering
    \includegraphics[width=1.04\textwidth]{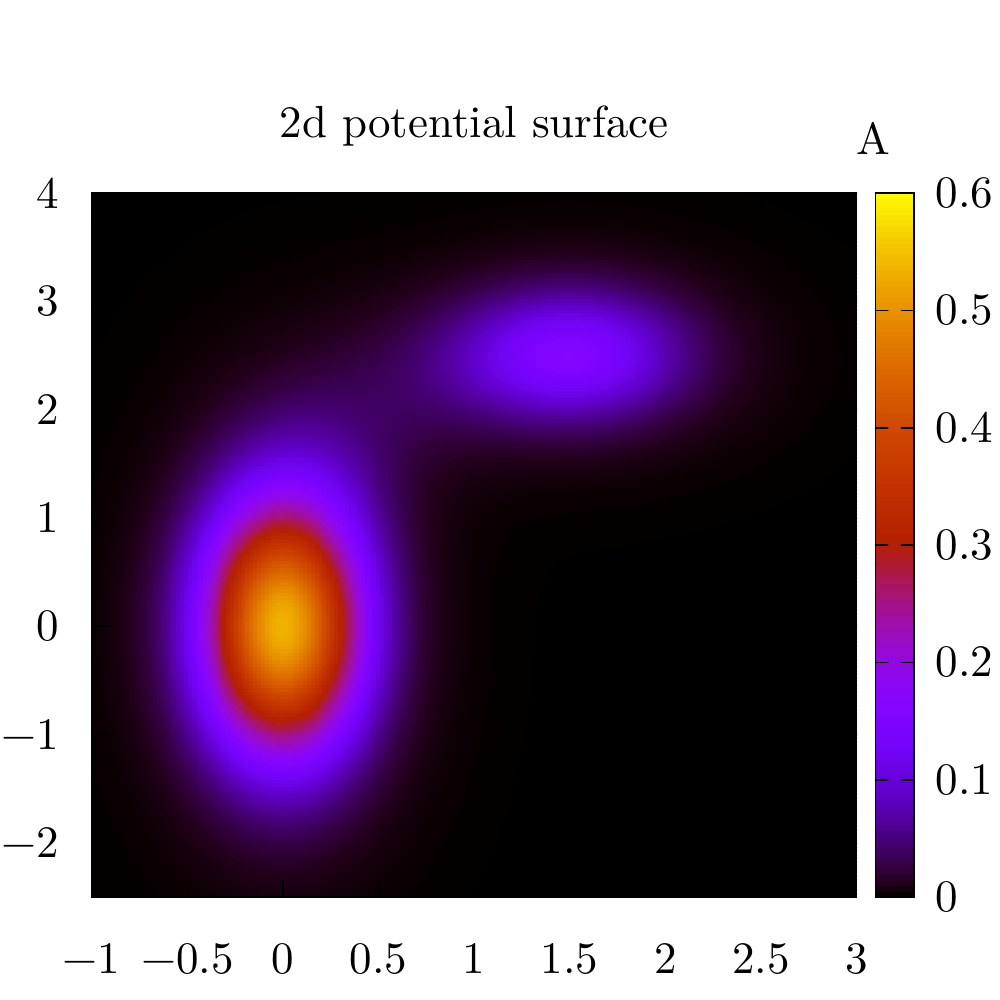}
  \end{subfigure}%
  
  \begin{subfigure}{\textwidth}
    \centering
    \includegraphics[width=.96\textwidth]{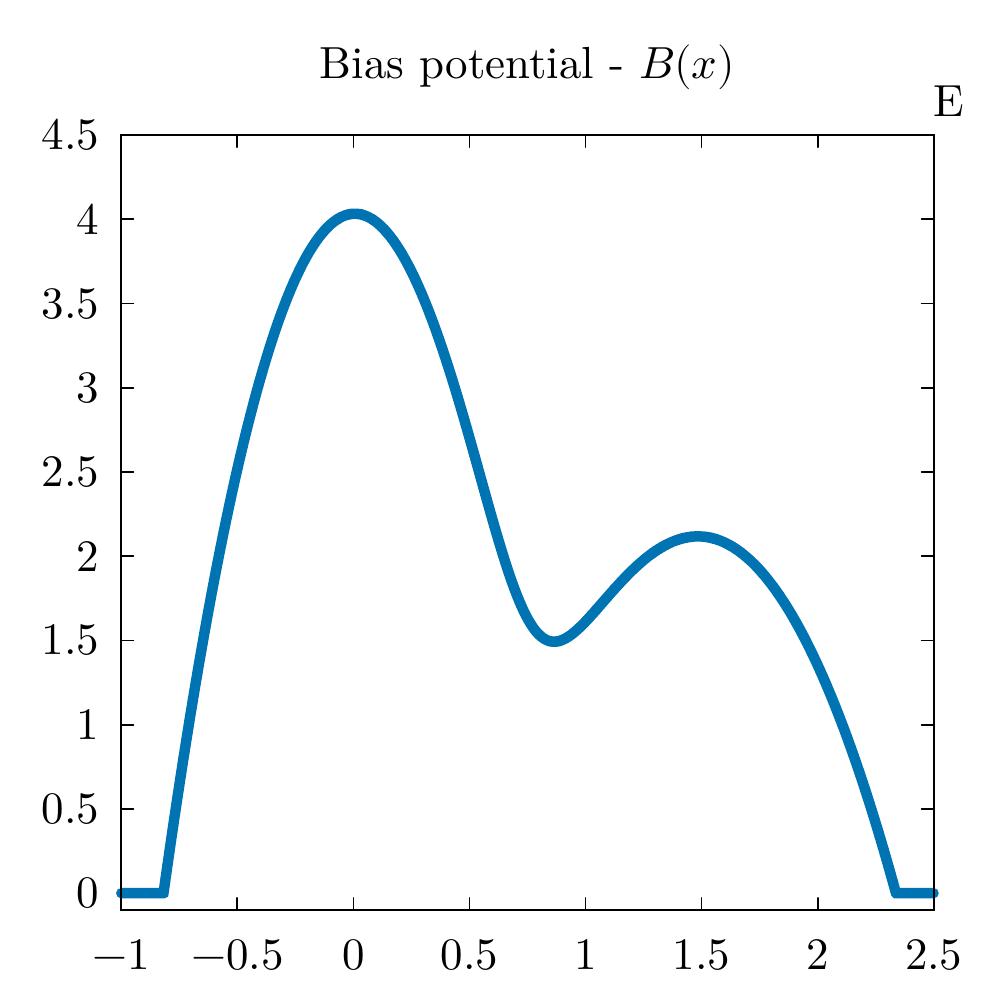}
  \end{subfigure}%
\end{minipage}
\hfill
\begin{minipage}{.72\textwidth}
  \begin{subfigure}{\textwidth}
    \includegraphics[width=\textwidth]{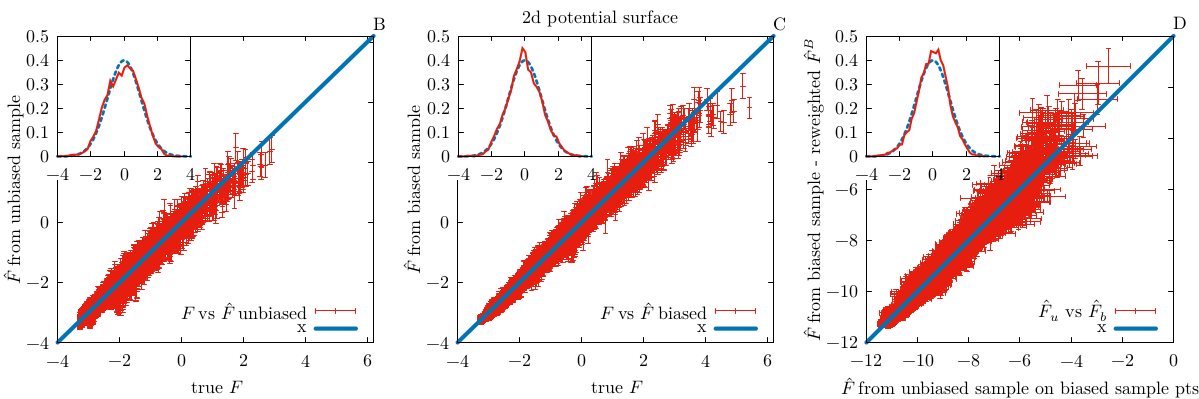}
  \end{subfigure}
  \begin{subfigure}{\textwidth}
    \includegraphics[width=\textwidth]{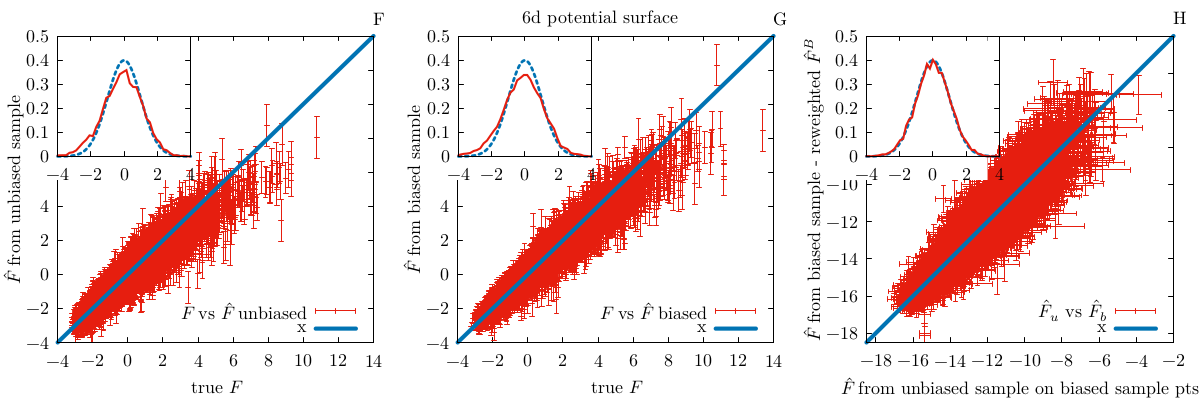}
  \end{subfigure}
\end{minipage}
    \caption{In the first column: (A) two-dimensional double-well potential surface used in simulations; (E) bias potential used along $x$ coordinate in the biased run for both analytic potentials. In the second and third column comparison of PA\textit{k} and bPA\textit{k} estimates against the analytic free energy showing correlation plots and pull distribution; (B),(C) respectively unbiased and biased case for 2d potential; (F),(G) respectively unbiased and biased case for 6d potential. (D),(H) Statistical test comparing bPA\textit{k} to PA\textit{k} on the points of the biased samples for 2d and 6d potentials respectively. All free energies and the bias potentials are in units of $k_BT$.}
    \label{fig:stattest_anal}
\end{figure*}

\subsection{Analytic conditions for punctual reweighting in Umbrella Sampling}
\label{ssec:methods-analitic_rw}
Looking at eq \ref{eq:F-rho_biased-explicit_dimred} we see that if there exists a map $\mathbf{\sigma} \mapsto \hat{\mathbf{s}}(\mathbf{\sigma})$ associating to each point $\mathbf{\sigma}(\mathbf{x})$ a unique value $\mathbf{s}(\mathbf{x})=\hat{\mathbf{s}}(\mathbf{\sigma}(\mathbf{x}))$ then $B(\mathbf{s}(\mathbf{x}))$ can be formally expressed as an explicit function of $\mathbf{\sigma}(\mathbf{x})$ and the exponential factor $e^{\, \beta  B(\hat{\mathbf{s}}(\mathbf{\sigma}))}$ can be brought out of the integral. Hence 
eq \ref{eq:F-rho_biased-explicit_dimred} for $\tilde{\mathbf{\sigma}}=\mathbf{\sigma}(\mathbf{x}_i)$ takes the form
\begin{equation}
    F(\mathbf{\sigma}(\mathbf{x}_i)) =  F^B(\mathbf{\sigma}(\mathbf{x}_i)) - B(\mathbf{s}(\mathbf{x}_i)).
\label{eq:F_reweight-analytic}
\end{equation}
For future reference, we call the existence of such map the map-existence condition (MEC). A consequence of the MEC is that if for two configurations $\mathbf{x}_1,\mathbf{x}_2$ we have $\sigma(\mathbf{x}_1)=\sigma(\mathbf{x}_2)$ then one cannot have $\mathbf{s}(\mathbf{x}_1) \ne \mathbf{s}(\mathbf{x}_2)$. Again, we stress the MEC is required only for the configurations $\mathbf{x}$ in the thermal ensemble. In fact, in molecular systems the interactions among atoms strongly reduce the independent directions in which the system can move. For this reason the ensemble density $\rho(\mathbf{x})$ is almost vanishing on a big portion of $\mathbb{R}^N$. The simplest case in which the MEC is verified is for $\mathbf{s}(\mathbf{x}) = \mathbf{s}(\mathbf{\sigma}(\mathbf{x}))$, namely when $\mathbf{s}$ is an explicit functions of the coordinates $\mathbf{\sigma}$. In this case $\hat{\mathbf{s}} \equiv \mathbf{s}$. However, eq \ref{eq:F_reweight-analytic} can also be valid if $\rho^B$ is estimated on the $\mathbf{\sigma}$ but $\mathbf{s}$ is an explicit function of different coordinates $\sigma'$, as long as these can be expressed as function of $\sigma$. This is true if all relevant $\mathbf{\sigma}'$ can be parametrised by an explicit function $\mathbf{\sigma}'=\varphi(\mathbf{\sigma})$. In this case $\mathbf{s}(\mathbf{x}) \equiv \mathbf{s}(\mathbf{\sigma}'(\mathbf{x})) \equiv \mathbf{s}(\varphi(\mathbf{\sigma}(\mathbf{x})))$ and $\hat{\mathbf{s}} \equiv \mathbf{s} \circ \varphi$.

\subsection{Reweighting with a punctual estimator}
In the cases where eq \ref{eq:F_reweight-analytic} holds, if one is able to estimate $F^B(\mathbf{\sigma}(\mathbf{x}_i))$ via an unbiased punctual estimator $\hat{F}^B_i$, the unbiased free energy at point $i$ can be estimated as:
\begin{equation}
    \hat{F}_i  := \hat{F}^B_i-B_i
\label{eq:bPAk-i-estim}    
\end{equation}
where $B_i:=B(\mathbf{s}(\mathbf{x}_i))$ is simply the numerical value of the bias applied when generating datapoint $\mathbf{x}_i$. Eq \ref{eq:bPAk-i-estim} applies in principle to any punctual estimator of the biased free energy. By choosing a suitable $\hat{F}^B_i$, the meaning of $\hat{F}_i$ becomes operatively clear. Due to its properties, we propose to estimate $\hat{F}^B_i$ with PA\textit{k}, because it is the only non-parametric free energy estimator method to our knowledge that is punctual without taking the limit $\Delta \rightarrow 0$. With this specification, eq \ref{eq:bPAk-i-estim} defines a simple punctual reweighting scheme to estimate point by point the unbiased free energy of a set of data generated in a biased simulation. From now on we shall for brevity refer to this procedure as \textbf{bPA\textit{k}}.

We will show that the procedure defined in eq \ref{eq:bPAk-i-estim} gives consistent result even when the MEC is slightly violated, i.e. when $\mathbf{s}$ is not an explicit function of the $\mathbf{\sigma}$, but there exists a parametrisation $\varphi$ that, given some $\mathbf{\sigma}$, is able to capture most relevant features in the space of the $\mathbf{\sigma}'$s.

\subsection{Measuring the quality of a free energy estimate}
\label{ssec:methods-stat_test}
We now discuss how we validate the robustness of the punctual reweighting procedure that we called bPA\textit{k}. We compare its performance to that of PA\textit{k} on unbiased samples, since PA\textit{k} is already established as a good free energy estimator on multidimensional data. Thus, we choose our test systems such that we are able to generate both an unbiased and a biased equilibrium sample. In order to assess the statistical compatibility of the results obtained with PA\textit{k} and bPA\textit{k}, we compute the correlation plot between estimated and ground truth free energies and the distribution of the quantity known as \textit{pull} \cite{demortier2002everything}. The value of the pull between two observables $F^a$ and $F^b$ at point $i$ is given by:

\begin{equation}
    \chi_i := \frac{(F^a_i-F^b_i)}{\sqrt{\varepsilon_{F^a_i}^2 + \varepsilon_{F^b_i}^2}}
\end{equation}
where $\varepsilon_{F}$ indicates the uncertainty on the quantity $F$. If $F^a$ and $F^b$ are compatible, the distribution of the pull over a full statistical sample $\{\textbf{x}_i{\}}_i$ is expected to be a Gaussian with zero average and unitary variance: $\chi_i \sim \mathcal{N}(0,1)$. Using these tools we first of all compare the estimates of free energy in the case of PA\textit{k} and bPA\textit{k} directly to the true known value in the case of multidimensional analytic potentials that we sample numerically. Secondly, still in the analytic case, we compare directly the two estimators. Finally, we consider as realistic case MD simulations of a $9$-peptide; in this case there is no known ground truth free energy for the system, therefore the only sensible test is to directly compare the unbiased and the biased estimators.

\section{Results}

\subsection{Validation of bPA\textit{k} on data sampled from multidimensional analytic potential surfaces}\label{ssec:results-bPAk_vs_PAk}

As a first step, we test bPA\textit{k} on systems for which the ground truth potential is known analytically. We consider two functional forms: a double well potential in 2d, shown in Figure \ref{fig:stattest_anal}A, and a 6d potential which is exactly as the previous one in the first 2 dimensions plus 4 other convex (harmonic) directions. For both potential we sample 10.000 points from a biased and an unbiased 
simulation. In order to bias the dynamics we apply as bias potential the inverse of the analytic free energy along the $x$ axis, cutoffed such that it is non-zero only on a finite interval (Figure \ref{fig:stattest_anal}E).

We apply PA\textit{k} to the unbiased sample and bPA\textit{k} to the biased sample, obtaining two estimates of the free energies, which in these two cases should coincide with the analytic potential energy. We test the two estimators directly against the analytically known ground truth with the statistical test described in section \ref{ssec:methods-stat_test}

Looking first at the 2-dimensional case, Figures \ref{fig:stattest_anal}B and \ref{fig:stattest_anal}F, the normal distribution with unitary variance and zero mean seems in both cases well approximated; the correlation is good, with the difference that the biased simulation explores regions at higher free energy, as expected. Also in the 6-dimensional case, Figures \ref{fig:stattest_anal}C and \ref{fig:stattest_anal}G, the agreement with the ground truth potentials is good. While for PA\textit{k} (unbiased case) this had previously been shown, also for bPA\textit{k} we can conclude that the method  performs excellently, at least in these two simple cases, and gives a statistically unbiased estimate of the correct free energies.

We now consider the case in which the ground truth unbiased free energy is not known explicitly, but one is able to generate a sample of data from the unbiased distribution. In this case, the unbiased and biased simulations sample different sets of points. Thus, since PA\textit{k} only gives a punctual estimate of the free energy, we need to define an interpolation scheme in order to directly compare PA\textit{k} and bPA\textit{k} results. For every point in the biased sample, we apply the PA\textit{k} procedure described in section \ref{ssec:PAk} adding only that point to the unbiased sample. To take into account the fact that this point does not actually belong to the sample we simply substitute $l \rightarrow l+1$ and set $\nu_{i,1}=0$ in eq \ref{eq:F_PAk}.

Figure \ref{fig:stattest_anal} (D),(H) shows the results of this test for the 2d and the 6d potentials; all samples generated have 10.000 points. Excellent compatibility between PA\textit{k} and bPA\textit{k} for both potentials is displayed. The example illustrates that the compatibility between a free energy estimated with and without the bias can be demonstrated also if the free energy is not known analytically. This will be used to analyse the simulation of the peptide.
\subsubsection{Comparison with standard reweighting on a finite neighbourhood}
\label{sssec:punctual_vs_standard_rw}
\begin{figure*}
    \centering
    \includegraphics[width=\textwidth]{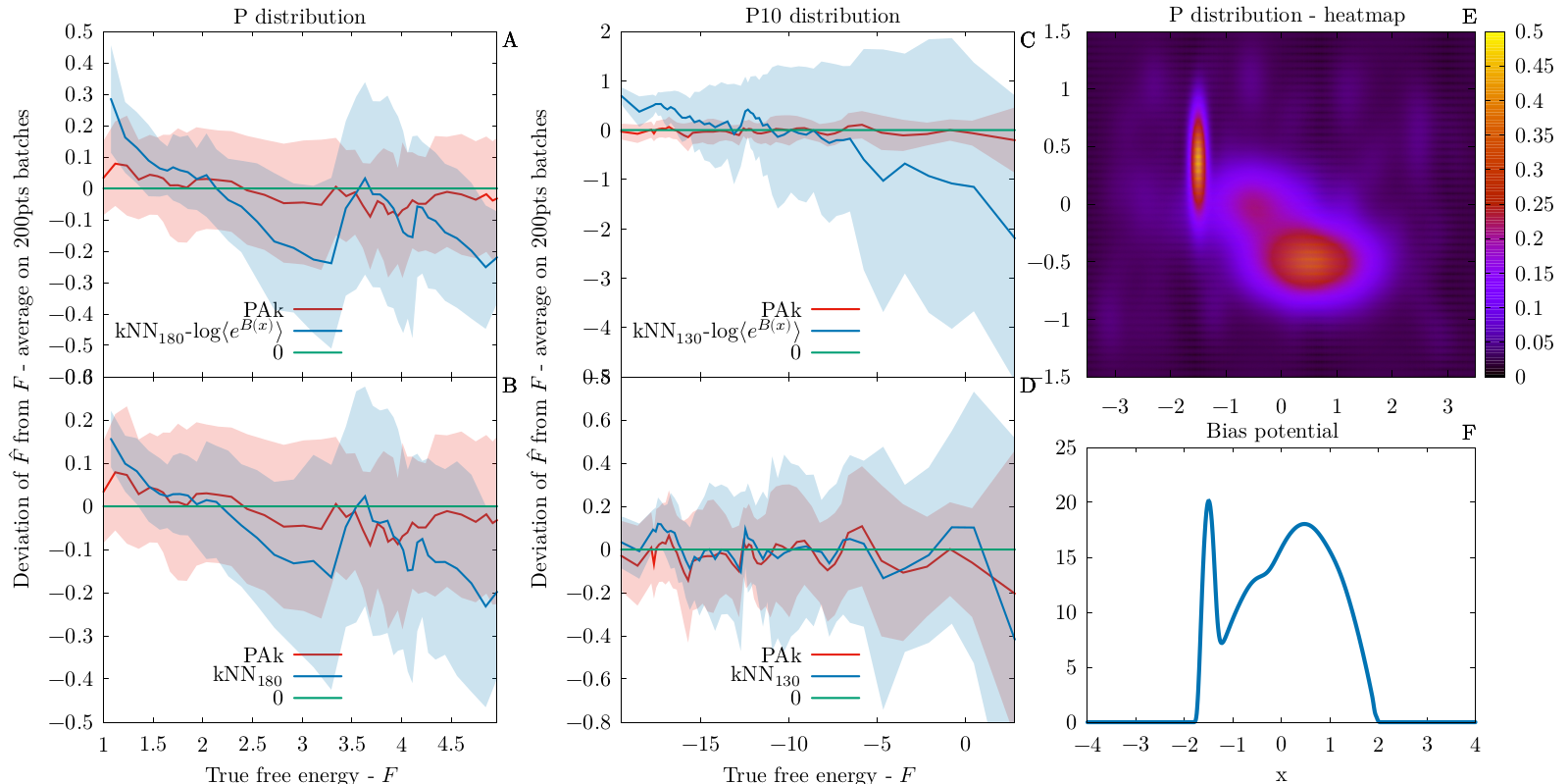}
\caption{Comparison of different biased free energy estimators and reweighting protocols on datasets sampled from two distributions: $p$ and $p^{10}$. (E) Heatmap representation of unbiased $p$. (F) Bias applied to $p^{10}$, which is exactly ten times the one applied to $p$. In both cased $d=D=2$, so what we call unbiased free energy corresponds identically to the potential energy entering the Boltzmann factor if $p$ and $p^{10}$ are interpreted as canonical p.d.f.'s. (A-D) . All estimates are performed on $10000$ data points. The values represented on the vertical axes are averages over batches of $200$ points. In solid transparent bars the sample standard deviation of the batch. (A),(B) refer to $p$;  (C),(D) refer to $p^{10}$. The chosen values of $180$ and $130$ for the \textit{k}-NN estimators are chosen as the average optimal value of \textit{k} predicted by PA\textit{k} for the datasets. (A),(C) bPA\textit{k} compared to \textit{k}-NN with standard reweighting. (B),(D) bPA\textit{k} compared to \textit{k}-NN, both with punctual reweighting}
\label{fig:glassy_pot}
\end{figure*}

We compare the performance of bPA\textit{k} with that of other kernel methods in the estimate of the free energy from biased data. Firstly, we compare the results of bPA\textit{k} to those of \textit{k}-NN reweighted in the standard way illustrated in eq \ref{eq:F_hat_bias-Kernel}, i.e. by subtracting to the estimated biased free energy the quantity $\log \langle e^{\, \beta B(\mathbf{s}_j)} \rangle_j$. Secondly, we apply the punctual reweighting also to \textit{k}-NN, for which this ansatz is not justified by a demonstration of punctuality of the density estimator, but becomes correct only in the limit $\Delta \rightarrow 0$.

We use two datasets in $2$ dimensions: one sampled from the probability density function $p$ represented in Figure \ref{fig:glassy_pot}E; the other one sampled from $p^{10}$ re-normalised to $1$, which for further reference we simply indicate by $p^{10}$. The two systems display metastability between the two main basins. By construction, the potential barrier in $p^{10}$ is exactly $10$ times as high as the one in $p$. The simulations are biased along the $x$ coordinate, with a bias obtained from the histogram along $x$ of the unbiased sampling (see Figure \ref{fig:glassy_pot}F). 

For both distributions, bPA\textit{k} drastically outperforms the \textit{k}-NN method reweighted in a standard manner (Figures \ref{fig:glassy_pot}A,\ref{fig:glassy_pot}C). With punctual reweighting, the quality of \textit{k}-NN estimates of the unbiased free energy improve. However, bPA\textit{k} still shows a better performance along the whole range of free energy values, especially for high values (Figures \ref{fig:glassy_pot}C,\ref{fig:glassy_pot}D). Furthermore, bPA\textit{k} yields much better relults also looking at the pull distributions between the estimated and the ground truth free energies in all reweighting schemes, a sign that bPA\textit{k} estimator is preferable to standard \textit{k}-NN also in terms of error estimate.

\subsection{Application of bPAk on an all-atom based simulation of a peptide}
\begin{figure*}[!ht]
    \centering
    \includegraphics[width=\textwidth]{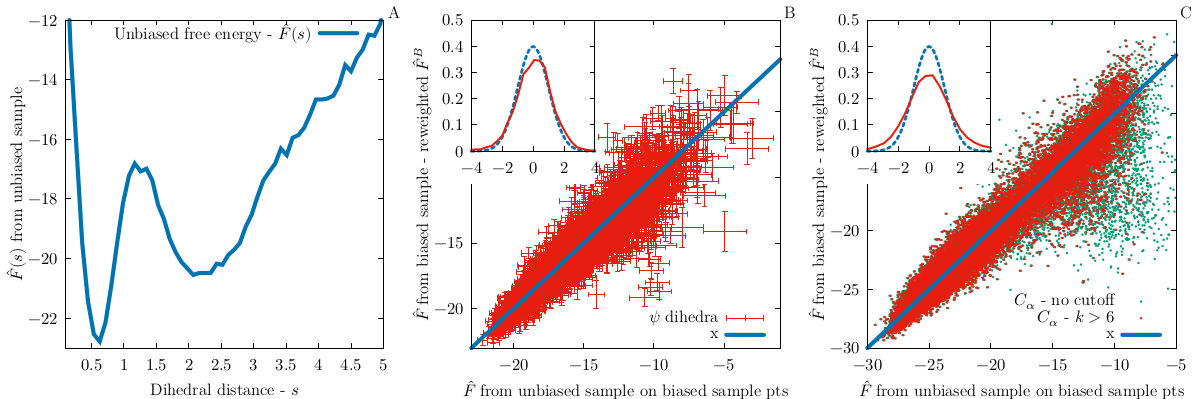}
\caption{(A) Free energy of the CLN025 peptide computed as histogram of the collective variable $s$. 
(B),(C) Statistical test comparing bPA\textit{k} to PA\textit{k} on the points of the biased sample in the case of the $\psi$-dihedral angles and of the $C_{\alpha}$ distances respectively. In (C) the error bars have been omitted from the correlation plot for a better readability; green dots represent all the points in the biased dataset; the red dots neglect all points with $\hat{k}_i \leq 6$.}
\label{fig:CLN025}
\end{figure*}
In order to test our method on a realistic system we choose a $\beta$-hairpin called CLN025\cite{honda2008crystal} 
. This molecule is a small protein of 10 residues and 166 atoms and is one of the smallest peptides that display a stable secondary structure, in this case a $\beta$-sheet. Thanks to the relatively small size of the molecule we are able to produce both a long unbiased MD trajectory and a biased one. In this way we are able to estimate the ground truth free energy of the system computing PA\textit{k} on the unbiased trajectory and to compare it with bPA\textit{k} results. 

We simulate the protein in Gromacs in explicit solvent. Since we are not interested in the precise phisical chemistry of the system, we use quite a small box with 931 water molecules. To enhance the sampling of configuration space, we run a Replica Exchange MD \cite{sugita1999replica} simulation with 16 replicas using equispaced temperatures from 340K to 470K as done previously in ref \cite{Rodriguez2011}.

In order to bias the trajectory, then, we choose as collective variable the $\psi$-dihedral distance from an equilibrium configuration, defined as:
\begin{equation}
    s = \sum_{n=1}^{9} \frac{1-cos(\psi_n - \psi_n^{ref})}{2}
    \label{eq:dihedral_distance}
\end{equation}
where $\psi_n$ denotes the $n$-th backbone $\psi$-dihedral angle of the peptide in the present configuration and $\psi_n^{ref}$ is the value of the same dihedral angle in a chosen reference equilibrium configuration (in our case we chose the crystal structure\cite{honda2008crystal}); this CV takes values from $s=0$, in the reference configuration, to $s=9$. We evaluate $s$ along the trajectory and compute the free energy $F(s)$ from a histogram (Figure \ref{fig:CLN025}A). We fit the lowest part ($\sim 10\;kJ mol^{-1}$) of the free energy with a sum of Gaussians $\tilde{F}(s) + c$ and use the negative of such sum as bias potential $B(s) = - \tilde{F}(s)$. Using PLUMED\cite{bonomi2009plumed} we run a umbrella sampling biased REMD simulation in the same setting of the unbiased one. 

Once the two trajectories are produced, we analyse them in the $9$-dimensional $\psi$-backbone-dihedra space. This choice implies of course a drastic dimensional reduction on the over-$400$-dimensional original atomic configuration space; still, even after this huge projection the system will show complex features and a reasonably high dimensionality, so that we are entitled to consider it a realistic case. Thus $D=9$ is the embedding space dimension. The distance between two configurations $\mathbf{X}^a$ and $\mathbf{X}^b$ in this space is:
\begin{equation}
    \theta(\mathbf{X}^a, \mathbf{X}^b) = \sqrt{ \sum_n (( \psi^a_n - \psi^b_n ))^2}
\end{equation}
where $(( \bullet ))$ stands for $2 \pi$-periodicity within the brackets.

We extract 9.500 points from the unbiased trajectory to use them as reference sample sample for PA\textit{k} Interpolator and 3.700 points from the biased one to be used as test sample (i.e. input of bPA\textit{k} and virtual points for PA\textit{k} Interpolator). The intrinsic dimension of the dataset is $d \sim 7$. The comparison between bPA\textit{k} estimate and our ground truth free energy (output of PA\textit{k} Interpolator), in Figure \ref{fig:CLN025}B, shows excellent agreement, despite the high dimensionality.

\subsubsection{Robustness of bPA\textit{k} under change of metric}
\label{sssec:change_of_metric}

We finally test the robustness of bPAk using a different coordinate system, in which $\mathbf{s}$ is not anymore an explicit function of the $\mathbf{\sigma}$. We choose as coordinates the distances among alpha carbons ($C_{\alpha}$) distances. Since we have 10 residues there are $10$ alpha carbons, thus $10 \times 9 / 2 = 45 $ pairwise distances between them; therefore the embedding dimension of the chosen space is now $D=45$. The metric on this space is the RMSD:
\begin{equation}
  d_{RMS}(\mathbf{X}^a, \mathbf{X}^b) = \sqrt{\frac{2}{N(N-1)} \sum_{i > j} ( d^a_{i j} - d^b_{i j} )^2}
\end{equation}
where $\mathbf{X}^a$ and $\mathbf{X}^b$ are two configurations in the $45$-dimensional space and $d^q_{i j}$ is the distance between the $i$-th and the $j$-th $C_{\alpha}$ in configuration $\mathbf{X}^q$.

This time we take 37.000 reference points from the biased sample and 38.000 from the unbiased one. The estimated intrinsic dimension is $d \sim 9$. We carry out our protocol and show the result of the statistical test in Figure \ref{fig:CLN025}C, green dots. Looking at the correlation plot we see a divergence from linearity especially at higher values of the free energy. We know that such values are associated to low densities in configuration space, corresponding either to low $\hat{k}_i$ or to high values for the distance of the $\hat{k}_i$-th neighbour. We notice (Figure \ref{fig:CLN025}C, red dots) that neglecting all points with $\hat{k}_i \leq 6$, corresponding to $\sim 10\%$ of the data, the correlation plot improves sensibly. Our explaination is that $C_{\alpha}$-distance and $\psi$-dihedra metrics are equivalent in the sense defined in section \ref{ssec:methods-analitic_rw} for low free energy configurations, while the MEC is partly violated at high F.

\subsubsection{Cluster analysis}

\begin{figure*}[h]
    \centering
    \includegraphics[width=\textwidth]{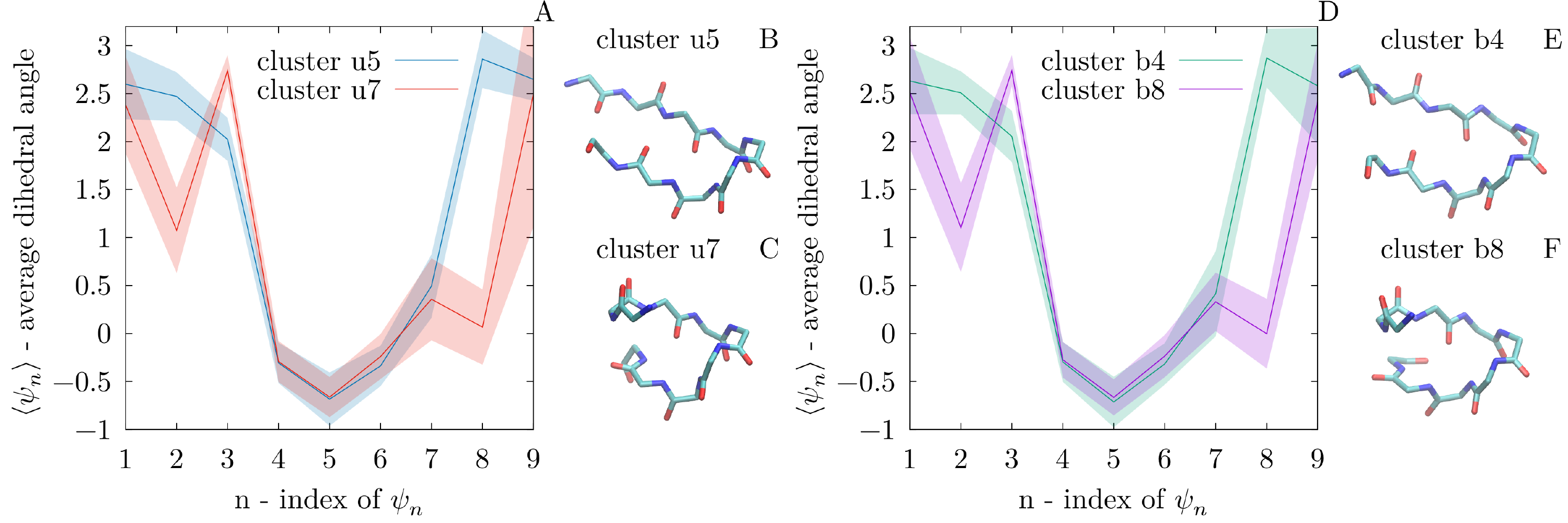}
\caption{\label{fig:cluster_analysis} Comparison between the two main clusters of the unbiased and of the biased trajectories. (A),(D) Average dihedral angle for each of the backbone dihedra for each cluster. (B),(C),(E),(F) backbone visualisation of the configurations closest in dihedral distance to the cluster average.}
\end{figure*}

As a last test to assess whether bPA\textit{k} captures the correct features of the free energy landscape we compare a cluster analysis in PA\textit{k} and bPA\textit{k}. As a clustering method we use Density Peak\cite{Rodriguez2014},\cite{DErrico2018}. In both cases we found eight clusters, but the two main clusters contain together more than $70\%$ of all the configurations. These clusters contain the most structured configurations, closest to the native state; they correspond in fact to the leftmost basin in the free energy of Figure \ref{fig:CLN025}A. Looking at Figure \ref{fig:cluster_analysis} we see both quantitatively and qualitatively that cluster u$5$ almost perfectly matches cluster b$4$ and the same happens with u$7$ and b$8$. Both couples of clusters present a structured $\alpha$-helix turn in the central region of the peptide (dihedra $4$-$6$); in the case of u$5$ and b$4$ the tails (dihedra $1$-$3$ and $8$-$9$) lie in the $\beta$-domain, hence the $\beta$-hairpin is fully folded; in clusters u$7$ and b$8$ the tails are unstructured; as expected, the first couple is also energetically slightly favoured. As for our initial purpose, we can conclude that bPA\textit{k} preserves also the cluster structure of the free energy.

\section{Conclusions}

We have presented bPA\textit{k}, a procedure to estimate the free energy in high-dimensional spaces starting from a sample of points generated in a biased simulation. The approach is based on a recently-introduced free-energy estimator called PA\textit{k}. 

PA\textit{k} is an optimised point-adaptive \textit{k}-NN algorithm, non-parametric and unsupervised. It takes as input a set of data expressed in terms of $D$ coordinates $\mathbf{\sigma}$ and produces punctual estimates of the free energy in these points. These coordinates can even coincide with the coordinates of the full configuration space. Importantly, we assume that these points lie on a manifold, embedded in the space of the $\mathbf{\sigma}$, of intrinsic dimensionality $d \leq D$. The densities in each point are computed in this low-dimensional manifold. However, the metric on the manifold is locally approximated as Euclidean, so volumes are measured in $\mathbb{R}^d$ without ever explicitly parametrising the embedded manifold. Hence, there is no explicit dimensional reduction in the space of the $\mathbf{\sigma}$, although this is done implicitly working in $d$ dimensions. At this level of description, the competitive advantage of PA\textit{k} is twofold: on one hand it optimally selects for every point in the dataset the size of the neighbourhood considered in the free energy estimate; on the other hand, the likelihood maximisation peculiar of PA\textit{k} extrapolates the value of the free energy in the limit of neighbourhood size going to zero, which makes the estimate more punctual than in other kernel-based methods.

The bPA\textit{k} protocol consists of computing the biased free energy at all points in the dataset using PA\textit{k} and then reweighting this quantity point by point simply subtracting the numerical value of the applied bias to retrieve an unbiased estimate. The simple additive form of this reweighting procedure is a nontrivial result. First of all, it crucially relies on the punctuality of PA\textit{k}. Secondly, since it involves integration over degrees of freedom which are not necessarily explicitly transversal to those involved in the computation of the bias potential, some reasonable but necessary requirements must be satisfied. We have described the condition (MEC) under which it is possible to reweight in an Umbrella-Sampling fashion the biased free energy over some coordinates $\mathbf{\sigma}(\mathbf{x})$ when the applied bias potential is a function of some possibly different CVs $\mathbf{s}(\mathbf{x})$. In short, this is possible if all the information necessary to define the biasing CVs $\mathbf{s}$ is encoded in the coordinates $\mathbf{\sigma}$ over which the free energy is computed. In other words, if the intrinsic manifold the $\{\mathbf{s}_i{\}}_i$ lie on can be mapped to a submanifold of the manifold the $\{\mathbf{\sigma}_i{\}}_i$ lie on. A case in which the MEC is violated on a small subset of configurations has been presented in the case of the CLN025 peptide. The derivation of the MEC and the consequent punctual form of the reweighting is valid in general for all punctual estimators, i.e. those for which the kernel introduced in eq \ref{eq:F_bias-Kernel} approaches a delta function. In order to apply the same reasoning to other finite-size-kernel methods, such as \textit{k}-NN, multidimensional histograms, gaussian KDE or other than involve the partitioning of the space of the CVs, one should require that the value of the bias potential does not vary much over the neighbourhoods that such density estimates consider for each point. 

We have tested bPA\textit{k} comparing its results to various unbiased ground truth free energies, some known analytically, some estimated with PA\textit{k} from an unbiased simulation. In all tested cases the pull distribution proved bPA\textit{k} to be an unbiased estimator of the ground truth values. In the case of two analytically-known distributions, we have also compared the performance of bPA\textit{k} to that of other estimators, both with standard and with punctual reweighting. While bPA\textit{k} is confirmed to be our best choice, punctual reweighting visibly improved the estimates also in the case of finite-size-kernel estimators. The opportunity of adopting punctual reweighting even with non-punctual estimators could be further investigated, but this is beyond the scope of this work.

\section*{Acknowledgements}
We thank Alex Rodriguez and Aldo Glielmo for several important suggestions and fruitful discussions.

\end{document}